\documentclass[showpacs,preprintnumbers,amsmath,amssymb,titlepage]{revtex4}
\usepackage{graphicx}
\usepackage{dcolumn}
\usepackage{bm}

\begin{document}

\title{Cooperative coupling of ultracold atoms and surface plasmons}

\author{Christian Stehle}
\affiliation{Physikalisches Institut and Center for Collective Quantum Phenomena in LISA+, Universit\"at
T\"ubingen, Auf der Morgenstelle 14, D-72076 T\"ubingen, Germany}
\author{Claus Zimmermann}
\affiliation{Physikalisches Institut and Center for Collective Quantum Phenomena in LISA+, Universit\"at
T\"ubingen, Auf der Morgenstelle 14, D-72076 T\"ubingen, Germany}
\author{Sebastian Slama}
\email{sebastian.slama@uni-tuebingen.de}
\affiliation{Physikalisches Institut and Center for Collective Quantum Phenomena in LISA+, Universit\"at
T\"ubingen, Auf der Morgenstelle 14, D-72076 T\"ubingen, Germany}

\date{\today}

\begin{abstract}
Cooperative coupling between optical emitters and light fields is one of the outstanding goals in quantum technology. It is both fundamentally interesting for the extraordinary radiation properties of the participating emitters and has many potential applications in photonics. While this goal has been achieved using high-finesse optical cavities, cavity-free approaches that are broadband and easy to build have attracted much attention recently. Here we demonstrate cooperative coupling of ultracold atoms with surface plasmons propagating on a plane gold surface. While the atoms are moving towards the surface they are excited by an external laser pulse. Excited surface plasmons are detected via leakage radiation into the substrate of the gold layer. A maximum Purcell factor of $\eta_\mathrm{P}=4.9$ is reached at an optimum distance of $z=250~\mathrm{nm}$ from the surface. The coupling leads to the observation of a Fano-like resonance in the spectrum. 
\end{abstract}
\maketitle

It is a long-sought goal of photonics to gain ultimate control over light-matter interactions on the single photon level \cite{Haroche06}. This goal is mainly motivated by the prospect of many possible applications, e.g. generation of single photons on-demand \cite{Imamoglu94, Kim99, He13}, control of the emission rate of quantum emitters \cite{Englund05}, nonlinearities on the single photon level \cite{Yamamoto99, Birnbaum05, Peyronel12} and the construction of single photon switches \cite{Volz12} and transistors \cite{Chang07, Neumeier13}. All of these applications require high cooperativity, a regime of cavity quantum electrodynamics (cqed) in which an optical emitter radiates with high probability into a distinct light mode \cite{Kimble98}. This regime is characterized by the condition that the cooperativity parameter 
\begin{equation}
\eta=4g^2/\left(\gamma\kappa\right)>1~,
\end{equation} 

with coupling constant $g$, natural line width of the emitter $\gamma$ and decay rate of the light field amplitude $\kappa$. Please note that this regime is contained in the strong coupling regime of cqed. The coupling constant $g$ is proportional to the dipole moment $d$ of the emitter and to the electric field $\cal{E}_\mathrm{1}$ connected to a single photon. High cooperativity can thus be reached by combining highly polarizable emitters with systems of large field enhancement. The latter condition can be reached e.g. in high-Q nanocavities with small mode volume~\cite{Khitrova06, Benson11}. An even stronger concentration of optical fields is possible with surface plasmons (SP) \cite{Barnes03, Gramotnev10, Schuller10}. While direct energy transfer from molecules to surface plasmons has been observed in early experiments \cite{Weber79, Pockrand80}, the interaction of single photon emitters and surface plasmons has attracted renewed interest in the last years \cite{Chang06, Tame13}. In this context modified emission rates were measured \cite{Amos97, Anger06, Andersen10}, single plasmons were excited \cite{Akimov07, Huck11}, and strong coupling between excitons and surface plasmons was demonstrated \cite{Bellessa04, Dintinger05, Vasa08, Hakala09, Gomez10, Schwartz11, Guebrou12, Tudela13}.\\

While in the above mentioned work artificial atoms (i.e. molecules, nanocrystals and quantum dots) are deposited as quantum emitters on the surface, proposals for trapping real ultracold atoms close to plasmonic structures have attracted much attention recently \cite{Chang09, Murphy09, Gullans12}. These proposals are based on the concept of dipole traps which are generated above plasmonic nanostructures, similar to the optical trapping of nanoobjects in plasmonically patterned light fields \cite{Righini07}. Real atoms have the advantage of being identical quantum emitters and having very narrow optical transitions with typical widths in the MHz range. Using quantum optics techniques clouds of atoms can be cooled to quantum degeneracy at temperatures on the order of nanokelvin \cite{Anderson95}. They can be trapped with ultrahigh precision in magnetic micotraps \cite{Fortagh07} and in optical dipole traps \cite{Grimm00} where they suffer very low intrinsic decoherence \cite{Wilk10}. Plasmonic traps might even further improve the control over the motion of atoms in the subwavelength regime, with dramatic consequences for interatomic scattering properties within one trap and atomic tunneling rates between neighboring traps \cite{Gullans12}. These traps could even be used for engineering strong p-wave interactions and the realization of exotic quantum many-body states with topological properties \cite{Diaz13}. Moreover, atoms which are positioned very close to plasmonic structures couple with high efficiency to surface plasmon modes which could be used for single photon applications and for enabling long-range interactions between atoms \cite{Chang09, Gullans12}. Please note that strong coupling is also pursued by combining cold atoms with nanophotonic waveguides \cite{Chang13, Thompson13, Goban14}.\\

\begin{figure}[ht]
	\includegraphics[]{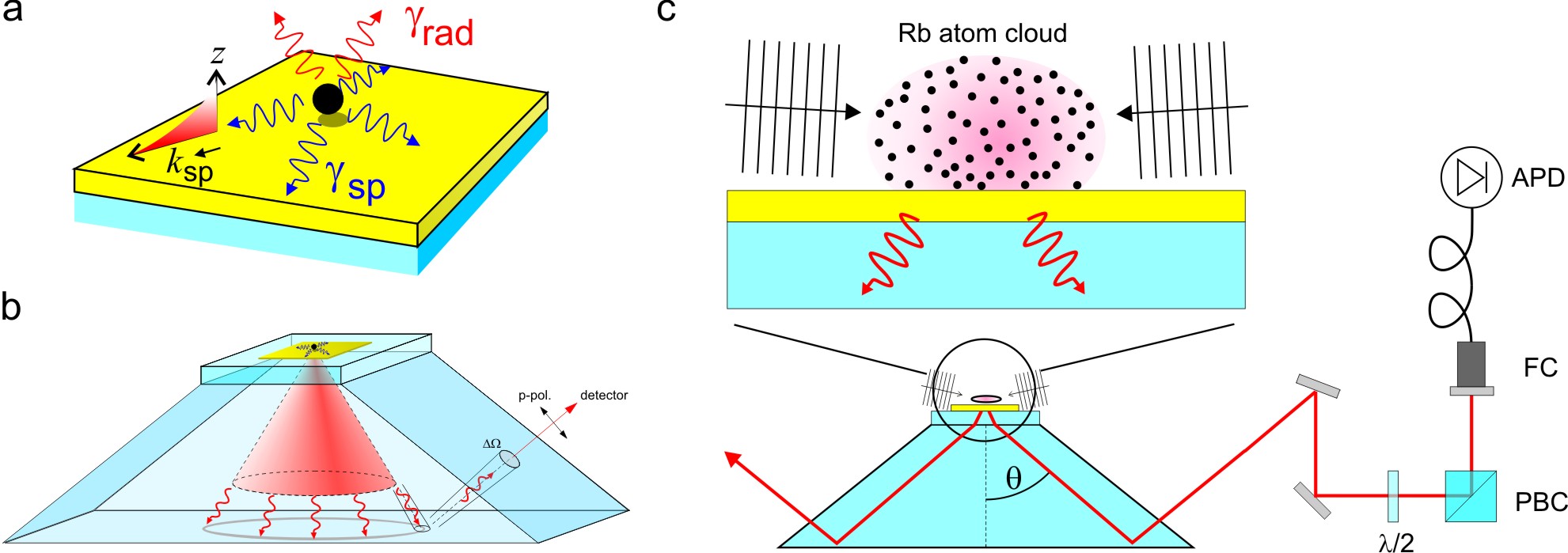}
	\caption{Schematic drawing of experimental situation. \textbf{a} An atom situated close to a metal surface emits photons radiatively with rate $\gamma_\mathrm{rad}$ into free space and nonradiatively with rate $\gamma_\mathrm{sp}$ into surface plasmons propagating on the metal surface with wavenumber $k_\mathrm{sp}$. \textbf{b} Surface plasmons on a thin metal film couple to the far field in the dielectric substrate via leakage radiation. The emitted light field is p-polarized. A detector collects photons under the solid angle $\Delta\Omega$. \textbf{c} In the experiment a cloud of ultracold Rubidium atoms is positioned very close to a gold film and is illuminated by a laser field. Photons which are emitted into the substrate are collected by an optical fibre coupler (FC) under an adjustable angle $\theta$ and are detected with an avalache photo diode (APD) with single photon sensitivity. The polarization of the detected light can be switched between s and p by a $\lambda /2$ waveplate and a polarizing beam cube (PBC).}
	\label{fig:setup}
\end{figure}

Despite the mentioned theoretical proposals only little experimental work has so far combined cold atoms with surface plasmons. In early work atoms have been reflected from evanescent light fields which were enhanced by surface plasmons on planar surfaces \cite{Esslinger93, Feron93, Schneble03}. Recently, we could show experimentally that potential landscapes for ultracold atoms can be tailored by plasmonic microstructures with the prospect of coupling single atoms to plasmonic devices \cite{Stehle11}. In this Article, we demonstrate direct and cooperative coupling of the fluorescence of cold atoms to surface plasmons propagating on a plane gold surface. The experiment is carried out with the experimental setup decribed in detail in~\cite{Stehle11}. An ultracold $^{87}$Rb atom cloud consisting of $N_\mathrm{at}\sim10^6$ atoms with a temperature of $T\sim 1~\mu\mathrm{K}$ is trapped in a magnetic trap inside a UHV chamber. The density of the cloud is given by 
\begin{equation}
n_\mathrm{at}(\vec r)=N_\mathrm{at}\left(8\pi^3\sigma_r^4\sigma_z^2\right)^{-{1/2}} e^{-(x^2+y^2)/2\sigma_r^2}e^{-(z-z_0)^2/2\sigma_z^2}~,
\end{equation}
with measured cloud width $\sigma_r\approx\sigma_z=76~\mu\mathrm{m}$. The maximum density in the center of the cloud is $n_\mathrm{at}^\mathrm{max}=1.4\times10^{17}~\mathrm{m}^{-3}$. By applying external magnetic fields the trapping position $z_0$ is moved towards a glass prism on which a sapphire substrate is attached (Fig. \ref{fig:setup}). Depending on the distance of the magnetic trapping  minimum from the surface attractive Casimir-Polder forces give rise to a potential barrier over which the atoms can move and are then accelerated towards the surface. Details on Casimir-Polder forces are contained in the Supplementary Method Section. The sapphire substrate carries square fields of gold layers with $\sim50~\mathrm{nm}$ thickness and  $\sim100~\mu\mathrm{m}$ side length. After the trapping minimum has been positioned in a variable distance from one of the gold fields the (moving) atoms are illuminated from the side with a $200~\mu\mathrm{s}$ long laser pulse. The laser intensity is adjusted to the saturation intensity $I_\mathrm{sat}=1.6~\mathrm{mW}/\mathrm{cm}^{2}$ corresponding to the $D2$-line of $^{87}$Rb. The incoming light field excites electric dipole oscillations in the atoms which are partially coupled to surface plasmons within the gold layer. The plasmon excitations decay into freely propagating photons~\cite{Hohenau11}. An avalanche photo diode (APD) with single photon detection efficiency measures the number of photons which is emitted under an adjustable angle $\theta$ into an aperture-limited opening angle of $\Delta\phi=\Delta\theta=0.6^\circ$. Due to Gaussian optics, this corresponds to a detectable spot size on the gold layer with radius $R_\mathrm{det}=32~\mu\mathrm{m}$. Above this spot, the atomic line density along the $z-$axis is given by $n_\mathrm{at}^\mathrm{max}\cdot\left(\pi R_\mathrm{det}^2\right)=46/100~\mathrm{nm}$. The detection scheme is sensitive to the polarization of the emitted light. This is crucial for discerning surface plasmon excitations from stray light~\cite{Raether88}.\\

\begin{figure}[ht]
	\includegraphics[]{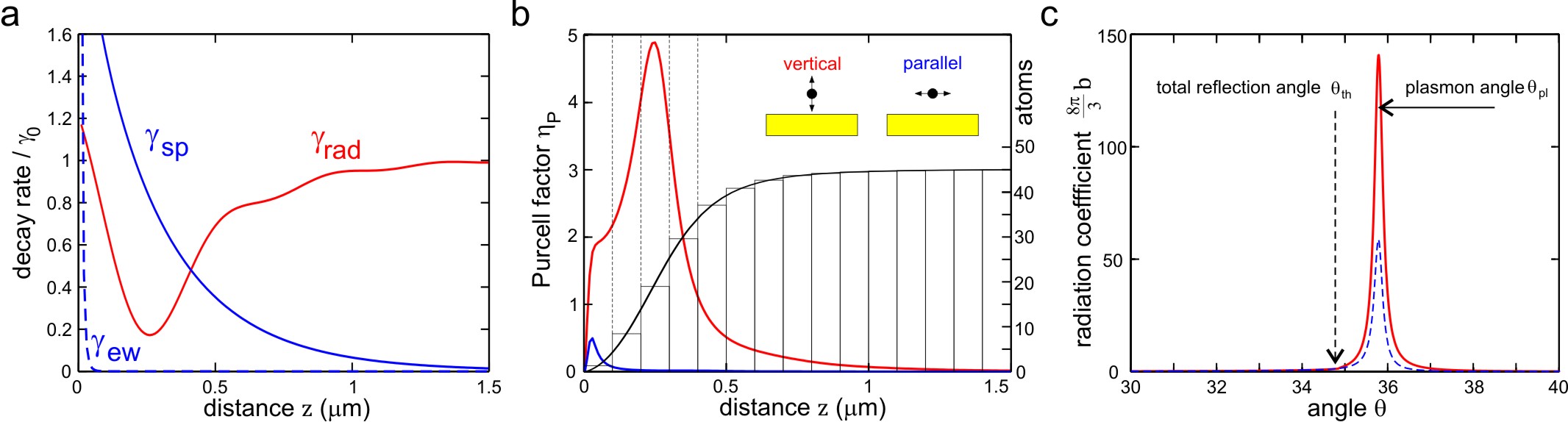}
	\caption{Radiation properties of a Rubidium atom at a gold surface. \textbf{a} Emission rates of a Rubidium atom with orthogonal dipole moment at a distance $z$ from a gold surface. Red line: Radiative emission into the far field ($\gamma_\mathrm{rad}$). Blue solid line: Nonradiative emission into surface plasmons ($\gamma_\mathrm{sp}$). Blue dashed line: Nanoradiative emission into evanescent waves ($\gamma_\mathrm{ew}$). All rates are normalized to the free space emission rate $\gamma_0$. \textbf{b} Purcell enhancement factor $\eta_\mathrm{P}=\frac{\gamma_\mathrm{sp}}{\gamma_\mathrm{rad}+\gamma_\mathrm{ew}}$ for a vertical dipole (red line) and a parallel dipole (blue line). The right axis corresponds to the black solid line and indicates the number of atoms in $100~\mathrm{nm}$ thick slices. \textbf{c} Radiation coefficient for emission into an angle $\theta$ within the substrate. The solid (dashed) line corresponds to a vertical dipole at a distance of $z=250~(500)~\mathrm{nm}$ from a $d=50~\mathrm{nm}$ thick gold layer on top of a sapphire substrate. The dielectric constants at the relevant wavelength of $\lambda=780~\mathrm{nm}$ are $\epsilon_1=-22.9+i\cdot1.42$ for gold and $\epsilon_2=3.0625$ for sapphire. Obviously, the surface plasmons are coupled into far field radiation in the sapphire substrate with high directionality under the plasmon angle $\theta_\textrm{pl}$. The angle of total reflection $\theta_\textrm{th}$ is that for an interface from sapphire to vacuum. Details on all the simulations are contained in the Supplementary Methods Section.}
	\label{fig:theory}
\end{figure}

Direct coupling between dipolar emitters and surface plasmons has been theoretically described in the context of nonradiative energy transfer \cite{Chance78, Sipe81}. The theories are based on a single electric dipole  $\vec{d}(t)$ oscillating with frequency $\omega$ at a position $\vec{r}_0$. It generates a radiation field $\vec{E}_0\left(\vec{r},t\right)$. The energy decate rate of the dipole is proportional to the product of the dipole moment and the electric field at the position of the dipole \cite{Sipe81}, 
\begin{equation}
\gamma=\frac{1}{2}\omega\cdot \mathrm{Im}\left[\vec{d}\cdot\vec{E}_0\left(\vec{r}_0\right)\right]~. 
\end{equation}
When the emitter is placed nearby a surface, also the part of the radiation field which is reflected at the boundary will interact with the dipole. This results in a modified decay rate, see Fig. \ref{fig:theory}~a). Moreover, also the emission pattern is changed. While in free space light can be emitted only radiatively, i.e. the wavenumber of the emitted light field is given by $k_0=\omega/c$, surfaces provide the possibility of nonradiative emission into bound surface waves. Their normal component of the wavevector is purely imaginary, whereas the parallel component has a value larger than that in free space, $k^\mathrm{sf}_\|>k_0$, i.e. bound surface waves propagate along the surface. Radiative $\gamma_\mathrm{rad}$ and nonradiative $\gamma_\mathrm{nrd}$ contributions to the total decay rate 
\begin{equation}
\gamma=\gamma_\mathrm{rad}+\gamma_\mathrm{nrd}
\end{equation}
can thus be separated by a Fourier expansion of $\gamma$ with respect to the dimensionless parameter $\kappa=k^\mathrm{sf}_\|/k_0$ and integration over the corresponding range $\kappa\leq1$ resp. $\kappa>1$~\cite{Sipe81}. Details on the different decay channels are contained in the Supplementary Methods Section. The theoretical results for a single $^{87}$Rb atoms at a distance $z$ from a gold surface are shown in Fig. \ref{fig:theory}~a. Here, nonradiative decay is mainly due to coupling to surface plasmon modes, with rate $\gamma_\mathrm{sp}$. These are collective electron oscillations in the metal surface connected with an electromagnetic wave propagating along the surface with wavenumber $k_\mathrm{sp}$. Decay into other bound modes like evanescent waves with rate $\gamma_\mathrm{ew}$ dominates only for very short distances $z\lesssim10~\mathrm{nm}$ \cite{Chance78}. Due to the symmetry of the problem the excited surface plasmon mode is cylindrically symmetric. For large distances from the atom it is well described by Hankel functions \cite{Arc09}, whereas for short distances where this approach fails, we describe the plasmon mode by the near field of a dipole close to a perfect conductor using the image dipole method. Details on the plasmon mode and its parameters are contained in the Supplementary Methods Section. Please note, that atomic collective effects can occur in dense ensembles of emitters which further influence the emission properties. In our setup atomic collective effects cannot be observed due to the low atomic density with a mean interatomic distance of $1.9~\mu\mathrm{m}$. Details on possible atomic collective effects are contained in the Supplementary Methods Section.\\

The efficiency of the direct coupling to surface plasmons is defined via the Purcell factor \cite{Chang07, Chang07b}
\begin{equation}
\eta_\mathrm{P}=\frac{\gamma_\mathrm{sp}}{\gamma_\mathrm{rad}+\gamma_\mathrm{ew}}~.
\end{equation}
Please note, that a large Purcell enhancement $\eta_\mathrm{P}>1$ is equivalent to the regime of high cooperativity in cqed \cite{Tanji11}. As shown in Fig. \ref{fig:theory}~b $^{87}$Rb atoms in a distance range of $20~\mathrm{nm}<z<410~\mathrm{nm}$ are coupled to plasmons in the gold surface with high cooperativity, if their atomic dipoles are oscillating vertically to the surface. In contrast, atoms with parallelly oscillating dipoles (with respect to the surface) do not reach the regime of high cooperativity. A pecularity of atoms as compared to molecules and quantum dots is the fact that the dipolar oscillation axis is not fixed by the orientation of the particle (the polarizability of an atom is radially symmetric), but it is parallel to the electric field of the exciting laser. Thus, the dipolar orientation can, in principle, be adjusted by the polarization of the incident laser pulse. In the present experiment the magnetic offset field of the trapping potential leads to a substantial Faraday rotation of the laser beam polarization while it propagates through the atom cloud. This reduces the number of vertically oscillating dipoles by $1/2$ and the plasmon excitation becomes independent of the pump laser polarization. The theory \cite{Sipe81} describes also the angular dependence of photons emitted into the substrate. The radiation coefficient for emission at an angle $\theta$ (see Fig. \ref{fig:setup} b) is given by
\begin{equation}
\label{eq:b}
b(\theta,z)=\frac{3}{8\pi}n_2\frac{k_x^2}{k_0^2}\left|T_{012}^p\frac{k_{z2}}{k_{z0}}\right|^2 e^{-2\mathrm{Im}(k_{z0})z}~,
\end{equation}
with parallel (to surface) wavevector $k_x$, free space wavevector $k_0=2\pi/\lambda$, and normal (to surface) wavevectors $k_{zj}$ in vacuum ($j=0$) and sapphire ($j=2$). The Fresnel field transmission coefficient $T_{012}^p$ is calculated for the considered three layer system (vacuum / $50~\mathrm{nm}$ gold layer / sapphire substrate) and incident parallel electric field oscillation. Details on the simulatoin are contained in the Supplementary Methods Section. The considered dipole is oscillating vertically at a distance $z$ from the surface. The radiation coefficient is normalized to the free space decay rate $\gamma_0$ such that the decay rate into a solid angle $\Delta\Omega$ of the substrate is given by
\begin{equation}
\label{eq:gamma2}
\gamma_2(\theta,\Delta\Omega,z)=\gamma_0\cdot\int b(\theta,z) d\Omega~.
\end{equation} 
Once excited, the plasmons emit photons highly directionally into the substrate under the so-called plasmon angle $\theta_\mathrm{pl}$, as illustrated in Fig. \ref{fig:theory}~c. Please note that due to the rotational symmetry of the plasmon mode photons are emitted on the surface of a cone, see Fig.~\ref{fig:setup}~b.\\

\begin{figure}[ht]
	\includegraphics[]{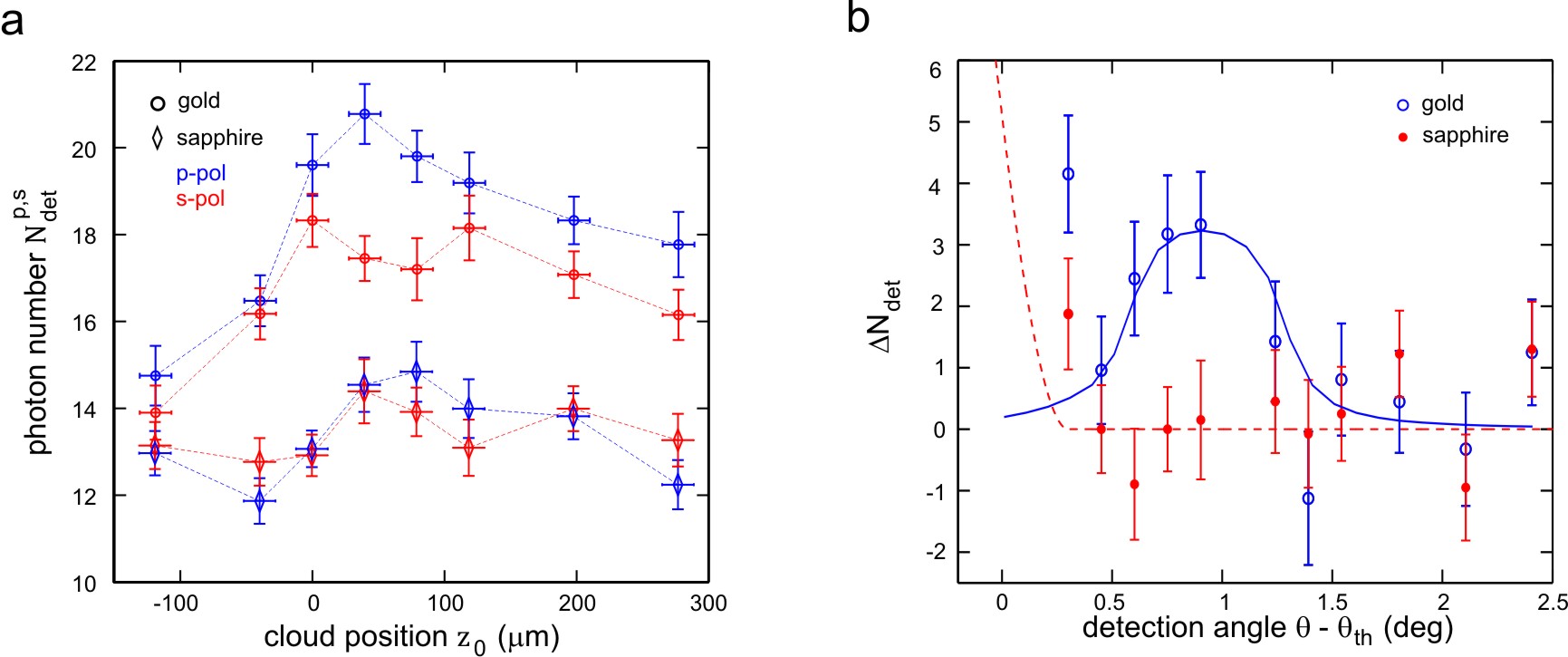}
	\caption{Measured photon numbers. \textbf{a} Number of detected p- and s-polarized photons at an angle of $\theta-\theta_\mathrm{th}=0.9^\circ$ above the gold surface and above the plain sapphire substrate. The position $z_0$ is the distance of the cloud center to the surface. \textbf{b} The photon number difference $\Delta N_\mathrm{det}=N_\mathrm{det}^\mathrm{p}-N_\mathrm{det}^\mathrm{s}$ at a distance of $z_0=41~\mu\mathrm{m}$ has a maximum at the plasmon angle above gold. The solid line is a simulation of nonradiative coupling of the atomic emission to the surface plasmon mode. The only fit parameter in this simulation is a loss factor $\eta_\mathrm{add}=0.36$ comprising the fibre coupling efficiency of the emitted light and additional losses at optical elements. Above the plain sapphire substrate no maximum is observed. The increase for small angles can be attributed to Fresnel transmission of light through surface roughness.}
	\label{fig:photons}
\end{figure}

The number of detected photons $N_\mathrm{det}^\mathrm{p,s}$ is plotted in Fig.~\ref{fig:photons}~a versus the trap center distance $z_0$ from the surface. As the atoms in the trap are moved towards the surface the number of detected photons is increasing. This general behavior is observed for both polarizations ($s$ and $p$) of the detected light and can be explained by the fact that more photons impinge on the surface when the atoms are closer. Negative values of $z_0$ correspond to trap centers which are inside the dielectric substrate for which the number of detected photons is decreasing. This decrease is caused by a loss of atoms from the trap as soon as they get into physical contact with the surface. The interesting feature in Fig.~\ref{fig:photons}a is observed at distances $0\lesssim z_0\lesssim 120~\mu\mathrm{m}$ which are comparable to the radial extension of the cloud such that a sufficient number of atoms can be found at submicron distance from the surface. At such distances an excess of p-polarized photons $N_\mathrm{det}^\mathrm{p}$ as compared to s-polarized photons $N_\mathrm{det}^\mathrm{s}$ is observed above gold. We attribute this excess to the excitation of surface plasmons. Thus, the following analysis concentrates on the observed photon difference 
\begin{equation}
\Delta N_\mathrm{det}=N_\mathrm{det}^\mathrm{p}-N_\mathrm{det}^\mathrm{s}~.
\end{equation} 
In Fig.~\ref{fig:photons}~b the measured angular dependence of $\Delta N_\mathrm{det}$ is shown comparing data which were obtained above the gold layer with data that were taken above the plain sapphire substrate. A maximum of $\Delta N_\mathrm{det}$ is observed at the plasmon angle  $\theta_\mathrm{pl}$ above gold. This is caused by the directive emission of surface plasmons into the substrate as calculated theoretically in Fig.~\ref{fig:theory}~c. We simulate this angular dependence by integration of Eq.~\ref{eq:gamma2}  
\begin{equation}
\label{eq:photnumber}
\Delta N_\mathrm{sim}=\int\left( n_\mathrm{at}(\vec r)\cdot n_\mathrm{rel}(z)\cdot\gamma_2(\theta,\Delta\Omega,z)\cdot\eta_\mathrm{ges}\cdot T_\mathrm{WW}\right)d^3r
\end{equation}
over the density of the atomic cloud $n_\mathrm{at}$ with interaction time $T_\mathrm{WW}$. The latter is given by the average time it takes for an atom with initial velocity zero to be detuned out of resonance due to radiation pressure. For $^{87}$Rb atoms illuminted with saturation intensity the interaction time is $T_\mathrm{WW}=84~\mu\mathrm{s}$ \cite{CohenTannoudji92}. The factor $n_\mathrm{rel}(z)$ describes the reduction of atomic density due to the accelerating influence of surface potentials, as determined in the Supplementary Methods. We integrate numerically in the radial direction over the detectable spot size with radius $R_\mathrm{det}$ and in the orthogonal direction from $z=0$ to $z=2~\mu\mathrm{m}$. We have checked that larger values for the upper integration limit do not change the result. The factor
\begin{equation}
\eta_\mathrm{ges}=\eta_\mathrm{qe}\cdot\eta_\mathrm{coh}\cdot\eta_\mathrm{ort}\cdot\eta_\mathrm{add}=0.015
\end{equation}
in Eq.~\ref{eq:photnumber} includes the quantum efficiency of the single photon counter $\eta_\mathrm{qe}=0.66$, the coherent scattering rate factor $\eta_\mathrm{coh}=\gamma_\mathrm{coh}/\gamma_0=0.125$ at saturation intensity, the fraction of orthogonally oscillating dipoles  $\eta_\mathrm{ort}=0.5$ due to Faraday rotation of the exciting laser field, and a factor $\eta_\mathrm{add}=0.36$ comprising any additional reduction of the overall detection efficiency, e.g. due to coupling of light into the optical multimode fibre and losses by optical elements. The obtained theoretical curve fits very well to the measured photon numbers in Fig.~\ref{fig:photons}~b. Another signature that emphasizes the role of surface plasmons is the fact that above sapphire, where plasmons do not exist, no peak is observed in $\Delta N_\mathrm{det}$. The additional increase of $\Delta N_\mathrm{det}$ above both substrates (gold and sapphire) at angles very close to the angle of total reflection $\theta_\mathrm{th}$ is attributed to the fact that part of the detected angle range $d \theta$ covers angles smaller than $\theta_\mathrm{th}$. Thus, some of the light which is emitted by the atoms can be transmitted directly through the surface. This behavior is confirmed by a simulation including the angular emission pattern of the atoms and Fresnel transmission into the substrate. Details on this simulation are contained in the Supplementary Methods Section. The simulation for sapphire is plotted as dashed line in Fig.~\ref{fig:photons}~b and explains the observation qualitatively. The quantitative deviation is attributed to surface roughness which is not taken into account in the simulations and which shifts the simulated curve to larger values of $\theta$.\\ 

\begin{figure}[ht]
	\includegraphics[]{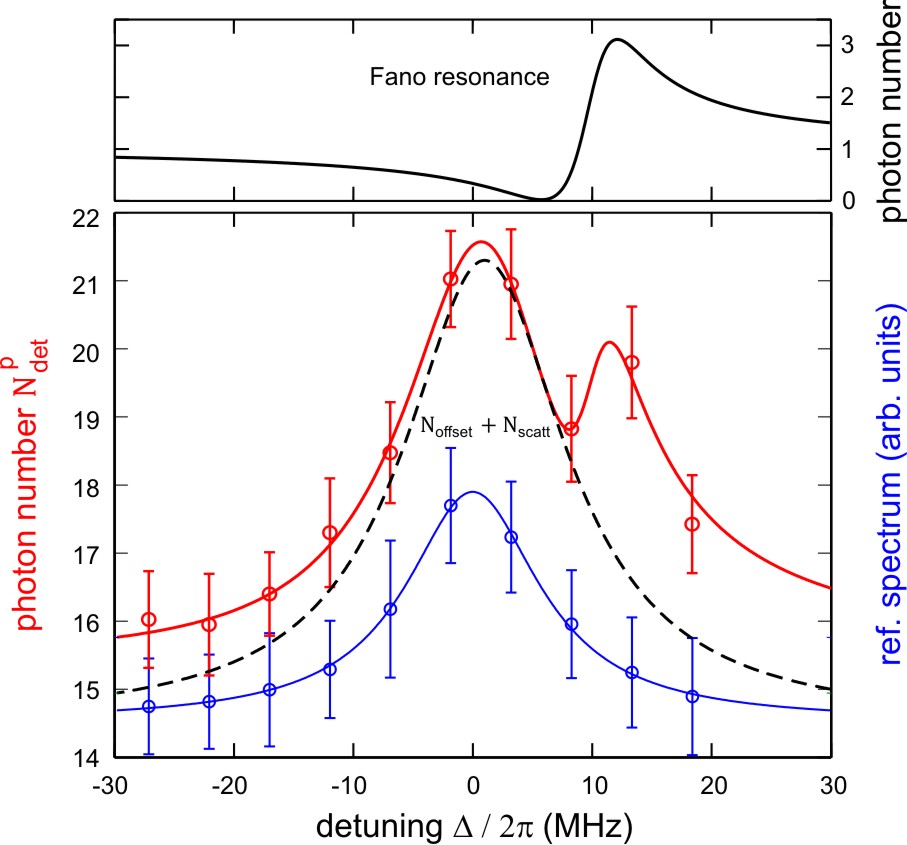}
		\caption{Measured spectrum. Red circles show the detected photon number $N_\mathrm{det}$ at an emission angle of $\theta-\theta_\mathrm{th}=0.9^\circ$ and cloud distance $z_0=41~\mu\mathrm{m}$. The data points are fitted by a model (solid line) including a constant offset $N_\mathrm{offset}$, a Lorentzian curve with amplitude $N_\mathrm{scatt}$ due to photons scattered by atoms far away from the surface and cooperatively coupled photons due to atoms very close to the surface that lead to a Fano profile. The dash-dotted line corresponds to the spectrum without the Fano contribution. For comparison, a reference spectrum (blue circles) was recorded simultaneously by saturation spectroscopy in a Rubidium vapor cell.}
	\label{fig:spectrum}
\end{figure}
 
Even more information on the details of the coupling can be obtained from the spectrum. Strong coupling can e.g. be identified by a splitting of the spectrum. Although the present experiment is far from the strong coupling regime, a dip in the measured spectrum is observed, see Fig \ref{fig:spectrum}. This dip can be attributed to a Fano resonance in the excitation of surface plasmons \cite{Fano61} as observed in a number of experiments in nanophotonics, see \cite{Fan14} and references therein. A Fano resonance is based on the coupling of a discrete state $|\chi\rangle$ via another discrete state $|\phi\rangle$ to a continuum of states $|k\rangle$ and the interference of this decay channel to the direct decay of $|\chi\rangle$ into $|k\rangle$ \cite{CohenTannoudji92}. The line profile of a Fano resonance is given by 
\begin{equation}
\label{eq:Fano1}
S_\mathrm{Fano}(\Delta)=\frac{\left|q_F+\frac{\Delta}{\gamma/2}\right|^2}{1+\left(\frac{\Delta}{\gamma/2}\right)^2}~,
\end{equation}
with decay rate $\gamma$ of state $|\phi\rangle$ and detuning $\Delta$. The Fano parameter $q_F$ is determined by the ratio of the resonant scattering amplitude (via the intermediate level) to the nonresonant scattering amplitude (direct channel). In our experiment the incident state is $|\chi\rangle=|N\rangle_{L}|g\rangle|0\rangle_{k_\mathrm{SP}}$, with $N$ photons in the pump laser field, the atom in the ground state $|g\rangle$ and zero photons in the surface plasmon mode with wavevector $k_\mathrm{SP}$. In the intermediate state, the atom has absorbed one photon from the pump laser field and has been excited into state $|e\rangle$, i.e. $|\phi\rangle=|N-1\rangle_{L}|e\rangle|0\rangle_{k_{SP}}$, and the final state $|f\rangle$ is the one where the photon has passed to the surface plasmon mode, i.e. $|f\rangle=|N-1\rangle_{L}|g\rangle|1\rangle_{k_{SP}}$. The final state can be reached whether (1) via the fluorescence of an atom (cooperative coupling) or (2) by direct excitation of the surface plasmon by the pump laser field. The latter process can occur due to surface roughness and is observed in Fig. \ref{fig:photons} a), where an excess  of p-polarized photons $\Delta N_\mathrm{direct}\approx 1$ is measured even when the atom cloud is still far away from the surface, resp. when it is lost due to surface interactions. This background excitation has to be compared with the situation when the atoms interact with the surface and excite surface plasmons by cooperative coupling, with $\Delta N_\mathrm{max}\approx 3$. The Fano parameter can thus be estimated by  
\begin{equation}
\label{eq:Fanoparameter}
q_F=\left(\frac{\Delta N_\mathrm{max}-\Delta N_\mathrm{direct}}{\Delta N_\mathrm{direct}}\right)^{1/2}\approx\sqrt{2}~,
\end{equation}
in which the square root is due to the fact that the observed photon numbers are given by the square of the scattering amplitudes. The spectrum in Fig. \ref{fig:spectrum} shows the number of detected photons while tuning the laser frequency across the Rubidium resonance. The measured signal is composed of (i) stray light (a constant offset), (ii) light scattered from atoms far away from the surface (a Lorentzian line shape) and (iii) excitation of surface plasmons (a Fano resonance). The measured data are thus fitted by the sum of the three contributions
\begin{equation}
\label{eq:Fano2}
N_\mathrm{det}^p(\Delta)=A_\mathrm{offset}+ A_{\text{scatt}}\frac{\gamma _{\text{scatt}}^{2}}{\gamma _{\text{scatt}}^{2}+4\left( \Delta -\delta _{\text{scatt}}\right) ^{2}}+ A_{\text{F}}\frac{\left|q_F+\frac{\Delta-\delta _{\text{F}}}{\gamma/2}\right|^2}{1+\left(\frac{\Delta-\delta _{\text{F}}}{\gamma/2}\right)^2}~.
\end{equation}
The best fit is obtained for amplitudes $A_\mathrm{offset}=14.5$, $A_{\text{scatt}}=6.8$, $A_{\text{F}}=1.0$, shifts of the resonance curves relative to a simultaneously recorded saturation spectrum of $\delta _{\text{scatt}}=2\pi\times1~\mathrm{MHz}$ and $\delta _{\text{F}}=2\pi\times10~\mathrm{MHz}$ and a broadening of the Lorentzian curve which is incorporated in $\gamma _{\text{scatt}}=2\pi\times16.5~\mathrm{MHz}$. Broadening and shifts can be explained by Zeeman-shifts of the transition frequencies due to magnetic trapping fields in the vacuum chamber on the order of ten Gauss. The width of the Fano resonance was held fixed at a value of $\gamma=\gamma_0=2\pi\times6~\mathrm{MHz}$. The Fano resonance itself is not broadened, because contributing atoms stem from a very thin layer above the surface.\\

Concluding, we have observed cooperative coupling between ultracold atoms and surface plasmons. Clouds of cold atoms are positioned in a magnetic trap at the surface of a sapphire substrate which is coated with a thin gold layer. From there the atoms move over a potential barrier towards the surface. They are illuminated and excited with a laser pulse and, depending on the distance of the atoms from the surface, the atomic excitation decays with high probability into surface plasmons that propagate along the metal surface. The plasmons are detected via photons that are emitted into the substrate. If the cloud is close enough to the surface more p-polarized than s-polarized photons are detected which is a signature of surface plasmons. The number of detected photons and the observed angular dependence of the emitted light fit quantitatively very well to the theoretical prediction with a maximum Purcell enhancement at a distance of $z_\mathrm{min}=250~\mathrm{nm}$ of $\eta_\mathrm{P}=4.9$. The recorded spectrum is asymmetric and can be explained by a Fano-like interference of surface plasmons excited via cooperative coupling and direct excitation of surface plasmons via surface roughness. Presently, the experiment is still far away from the strong coupling regime in which single atoms can perform full vacuum Rabi oscillations with plasmonic excitations. However, enhanced coupling may be reached by replacing the plane gold surface by a plasmonic nanostructure which acts as a cavity for surface plasmons. For a $\left(\lambda\times\lambda\right)-$metal square field ($\lambda=780~\mathrm{nm}$) with cavity mode volume of $V_\mathrm{mode}\sim\lambda^3$ the coupling constant is
\begin{equation}
\label{eq:couplingconstant}
g=\sqrt{3\pi\gamma_0c^3/\omega_0^2V_\mathrm{mode}}\sim2\pi\times10~\mathrm{GHz}~.
\end{equation}
A further enhancement of the coupling constant may by achieved by collective atomic effects, in which many atoms couple simultaneously to the same plasmon mode. While the atomic density in the present experiment is on the order of  $n_\mathrm{at}^\mathrm{max}\sim10^{17}~\mathrm{m}^{-3}$, typical densities in Bose-Einstein condensates on the order of $n_\mathrm{at}^\mathrm{max}\sim10^{20}~\mathrm{m}^{-3}$ are achievable. These results and perspectives demonstrate that hybrid systems consisting of cold atoms and surface plasmons have the potential for photonic devices on the single photon level with coupling constants hitherto unreached in cold atom experiments. Moreover, cooperative coupling of individual atoms with surface plasmons may lead to long-range interactions between different atoms for slightly increased atomic densities. The emission properties can then be more complex than for a single atom with quantum correlations arising between the atoms \cite{Choquette10}.


\begin{thebibliography}{10}
\bibitem{Haroche06}  Haroche, S. \& Raimond, J.-M. \textit{Exploring the quantum: Atoms, Cavities, Photons.} (Oxford Univ. Press, New York, 2006).
%
\bibitem{Imamoglu94}  Imamoglu, A. \& Yamamoto, Y. Turnstile device for heralded single photons: coulomb blockade of electron and hole tunneling in quantum confined p-i-n heterojunctions. \textit{Phys. Rev. Lett.} \textbf{72}, 210-213 (2011).
%
\bibitem{Kim99}  Kim, J., Benson, O., Kan, H. \& Yamamoto, Y. A Single-photon turnstile device. \textit{Nature} \textbf{397}, 500-503 (1999).
%
\bibitem{He13}  He, Y.-M. et al. On-demand semiconductor single-photon source with near-unity indistinguishability. \textit{Nature Nanotech.} \textbf{8}, 213-217 (2013).
%
\bibitem{Englund05}  Englund, D. et al. Controlling the spontaneous emission rate of single quantum in a two-dimensional photonic crystal. \textit{Phys. Rev. Lett.} \textbf{95}, 013904 (2005).
%
\bibitem{Yamamoto99}  Yamamoto, Y. \& Imamoglu, A. \textit{Mesoscopic Quantum Optics.} (Wiley \& Sons, 1999).
%
\bibitem{Birnbaum05}  Birnbaum, K.M. et al. Photon blockade in an optical cavity with one trapped atom. \textit{Nature} \textbf{436}, 87-90 (2005).
%
\bibitem{Peyronel12}  Peyronel, T. et al. Quantum nonlinear optics with single photons enabled by strongly interacting atoms. \textit{Nature} \textbf{488}, 57-60 (2012).
%
\bibitem{Volz12}  Volz, T. et al. Ultrafast all-optical switching by single photons. \textit{Nature Phot.} \textbf{6}, 605-609 (2012).
%
\bibitem{Chang07}  Chang, D.E., S\o rensen, A.S., Demler, E.A. \& Lukin, M.D. A single-photon transistor using nanoscale surface plasmons. \textit{Nature Phys.} \textbf{3}, 807 - 812 (2007).
%
\bibitem{Neumeier13}  Neumeier, L., Leib, M. \& Hartmann, M.J. et al. Single-Photon Transistor in Circuit Quantum Electrodynamics. \textit{Phys. Rev. Lett.} \textbf{111}, 063601 (2013).
%
\bibitem{Kimble98}  Kimble, H.J. Strong interactions of single atoms and photons in cavity qed. \textit{Physica Scripta T} \textbf{76}, 127 (1998).
%
\bibitem{Khitrova06}  Khitrova, G., Gibbs, H.M., Kira, M., Koch, S.W. \& Scherer, A. Vacuum Rabi splitting in semiconductors. \textit{Nature Phys.} \textbf{2}, 81-90 (2006).
%
\bibitem{Benson11}  Benson, O. Assembly of hybrid photonic architectures from nanophotonic constituents. \textit{Nature} \textbf{480}, 193-199 (2011).
%
\bibitem{Barnes03}  Barnes, W.L., Dereux, A. \& Ebbesen, T.W. Surface plasmon subwavelength optics. \textit{Nature} \textbf{424}, 824-830 (2003).
%
\bibitem{Gramotnev10}  Gramotnev, D.K. \& Bozhevolnyi, S.I. Plasmonics beyond the diffraction limit. \textit{Nature Photon.} \textbf{4}, 83-91 (2010).  
%
\bibitem{Schuller10}  Schuller, J.A. et al. Plasmonics for extreme light concentration and manipulation. \textit{Nature Mater.} \textbf{9}, 193-204 (2010).
%
\bibitem{Weber79}  Weber, W.H. \& Eagen, C.F. Energy transfer from an excited dye molecule to the surface plasmons of an adjacent metal. \textit{Opt. Lett.} \textbf{4}, 236-238 (1979).
%
\bibitem{Pockrand80}  Pockrand, I. \& Brillante, A. Nonradiative Decay Of Excited Molecules Near A Metal Surface. \textit{Chem. Phys. Lett.} \textbf{69}, 499-504 (1980).
%
\bibitem{Chang06}  Chang, D.E., S\o rensen, A.S., Hemmer, P.R. \& Lukin, M.D. Quantum Optics with Surface Plasmons. \textit{Phys. Rev. Lett.} \textbf{97}, 053002 (2006).
%
\bibitem{Tame13}  Tame, M.S., McEnery, K.R., Özdemir, S.K., Lee, J., Maier, S.A. \& Kim, M.S. Quantum Plasmonics. \textit{Nature Phys.} \textbf{9}, 329 - 340 (2013).
%
\bibitem{Amos97}  Amos, R.M., \& Barnes, W.L. Modification of the spontaneous emission rate of Eu$^{3+}$ ions close to a thin metal mirror. \textit{Phys. Rev. B} \textbf{55}, 7249-7254 (1997).
%
\bibitem{Anger06}  Anger, P., Bharadwaj, P. \& Novotny, L. Enhancement and Quenching of Single-Molecule Fluorescence. \textit{Phys. Rev. Lett.} \textbf{96}, 113002 (2006).
%
\bibitem{Andersen10} Andersen, M.L., Stobbe, S., S\o rensen, A.S. \& Lodahl, P. Strongly modified plasmons-matter interaction with mesoscopic quantum emitters \textit{Nature Phys.} \textbf{7}, 215 - 218 (2010).
%
\bibitem{Akimov07}  Akimov, A.V. et al. Generation of single optical plasmons in metallic nanowires coupled to quantum dots. \textit{Nature} \textbf{450}, 402 - 406 (2007).
%
\bibitem{Huck11}  Huck, A., Kumar, S., Shakoor, A. \& Andersen, U.L. Controlled Coupling of a Single Nitrogen-Vacancy Center to a Silver Nanowire. \textit{Phys. Rev. Lett.} \textbf{106}, 096801 (2011).
%
\bibitem{Bellessa04}  Bellessa, J., Bonnand, C. \& Plenet, J.C. Strong coupling between Surface Plasmons and Excitons in an Organic Semiconductor. \textit{Phys. Rev. Lett.} \textbf{93}, 036404 (2004).
%
\bibitem{Dintinger05}  Dintinger, J., Klein, S., Bustos, F., Barnes, W.L. \& Ebbesen, T.W. Strong coupling between surface plasmon-polaritons and organic molecules in subwavelength hole arrays. \textit{Phys. Rev. B} \textbf{71}, 035424 (2005).
%
\bibitem{Vasa08}  Vasa, P. et al. Coherent Exciton-Surface-Plasmon-Polariton Interaction in Hybrid Metal-Semiconductor Nanostructures. \textit{Phys. Rev. Lett.} \textbf{101}, 116801 (2008).
%
\bibitem{Hakala09}  Hakala, T.K. et al. Vacuum Rabi Splitting and Strong-Coupling Dynamics for Surface-Plasmon Polaritons and Rhodamine 6G Molecules. \textit{Phys. Rev. Lett.} \textbf{103}, 053602 (2009).
%
\bibitem{Gomez10}  G\'{o}mez, D.E. , Vernon, K.C., Mulvaney, P. \& Davis, T.J. Surface Plasmon Mediated Strong Exciton-Photon Coupling in Semiconductor Nanocrystals. \textit{Nano Lett.} \textbf{10}, 274-278 (2010).
%
\bibitem{Schwartz11}  Schwartz, T., Hutchison, J.A., Genet, C. \& Ebbesen, T.W. Reversible Switching of Ultrastrong Light-Molecule Coupling. \textit{Phys. Rev. Lett.} \textbf{106}, 196405 (2011).
%
\bibitem{Guebrou12}  Aberra Guebrou, S. et al. Coherent Emission from a Disordered Organic Semiconductor Induced by Strong Coupling with Surface Plasmons. \textit{Phys. Rev. Lett.} \textbf{108}, 066401 (2012).
%
\bibitem{Tudela13}  Gonz\'{a}lez-Tudela, A., Huidobro, P.A., Mart\'{i}n-Moreno, C. \& Garc\'{i}a-Vidal, F.J. Theory of Strong Coupling between Quantum Emitters and Propagating Surface Plasmons. \textit{Phys. Rev. Lett.} \textbf{110}, 126801 (2013).
%
\bibitem{Chang09}  Chang, D.E. et al. Trapping and Manipulation of Isolated Atoms Using Nanoscale Plasmonic Structures. \textit{Phys. Rev. Lett.} \textbf{103}, 123004 (2009).
%
\bibitem{Murphy09}  Murphy, B. \& Hau, L.V. Electro-Optical Nanotraps for Neutral Atoms. \textit{Phys. Rev. Lett.} \textbf{102}, 033003 (2009).
%
\bibitem{Gullans12}  Gullans, M. et al. Nanoplasmonic Lattices for Ultracold Atoms. \textit{Phys. Rev. Lett.} \textbf{109}, 235309 (2012).
%
\bibitem{Righini07} Righini, M., Zelenina, A.S., Girard, C. \& Quidant, R. Parallel and selective trapping in a patterned plasmonic landscape. \textit{Nature Phys.} \textbf{3}, 477-480 (2007).
%
\bibitem{Anderson95} Anderson, M.H., Ensher, J.R., Matthews, M.R., Wieman, C.E. \& Cornell, E.A. Observation of Bose-Einstein Condensation in a Dilute Atomic Vapor. \textit{Science} \textbf{269}, 198-201 (1995).
%
\bibitem{Fortagh07}  Fort\'agh, J. \& Zimmermann, C. Magnetic microtraps for ultracold atoms. \textit{Rev. Mod. Phys.} \textbf{79}, 235-289 (2007).
%
\bibitem{Grimm00}  Grimm, R., Weidem\"uller, M. \& Ovchinnikov, Y.B. Optical Dipole Traps for Neutral Atoms. \textit{Adv. At. Mol. Opt. Phys.} \textbf{42}, 95-170 (2000).
%
\bibitem{Wilk10}  Wilk, T. et al. Entanglement of Two Individual Neutral Atoms Using Rydberg Blockade. \textit{Phys. Rev. Lett.} \textbf{104}, 010502 (2010).
%
\bibitem{Diaz13}  Juli\'a-D\'iaz, B., Gra\ss , T., Dutta, O., Chang, D.E. \& Lewenstein, M. Engineering p-wave interactions in ultracold atoms using nanoplasmonic traps. \textit{Nature Commun.} \textbf{4}, 2046 (2013).
%
\bibitem{Chang13}  Chang, D.E., Cirac, J.I. \& Kimble, H.J. Self-Organization of Atoms along a Nanophotonic Waveguide. \textit{Phys. Rev. Lett.} \textbf{110}, 113606 (2013).
%
\bibitem{Thompson13} Thompson, J.D. et al. Coupling a Single Trapped Atom to a Nanoscale Optical Cavity. \textit{Science} \textbf{340}, 1202-1205 (2013).
%
\bibitem{Goban14} Goban, A. et al. Atom–light interactions in photonic crystals. \textit{Nature Commun.} \textbf{5}, 3808 (2014).
%
\bibitem{Esslinger93}  Esslinger, T., Weidem\"uller, M., Hemmerich, A. \& H\"ansch, T.W. Surface-plasmon mirror for atoms. \textit{Opt. Lett.} \textbf{18}, 450-452 (1993).
%
\bibitem{Feron93}  Feron, S. et al. Reflection of metastable neon atoms by a surface plasmon wave. \textit{Opt. Comm.} \textbf{102}, 83-88 (1993).
%
\bibitem{Schneble03}  Schneble, D., Hasuo, M., Anker, T., Pfau, T. \& Mlynek, J. Detection of cold metastable atoms at a surface. \textit{Rev. Sci. Instr.} \textbf{74}, 2685-2689 (2003).
%
\bibitem{Stehle11} Stehle, C. et al. Plasmonically tailored micropotentials for ultracold atoms. \textit{Nature Photon.} \textbf{5}, 494-498 (2011).
%
\bibitem{Hohenau11} Hohenau, A. et al. Surface plasmon leakage radiation microscopy at the diffraction limit. \textit{Opt. Exp.} \textbf{19}, 25749-25762 (2011).
%
\bibitem{Raether88}  Raether, H. \textit{Surface Plasmons on Smooth and Rough Surfaces and on Gratings} (Springer-Verlag, Berlin, 1988).
%
\bibitem{Chance78} Chance, R.R., Prock, A. \& Silbey. R. Molecular Fluorescence And Energy Transfer Near Interfaces. \textit{Adv. Chem. Phys.} \textbf{37}, 1-65 (1978).
%
\bibitem{Sipe81} Sipe, J.E. The Dipole Antenna Problem In Surface Physics: A New Approach. \textit{Surf. Science} \textbf{105}, 489-504 (1981).
%
\bibitem{Arc09}  Archambault, A., Teperik, T.V. , Marquier, F. \& Greffet, J.J. Surface plasmon Fourier optics. \textit{Phys. Rev. B} \textbf{79}, 195414 (2009).
%
\bibitem{Chang07b}  Chang, D.E., S\o rensen, A.S., Hemmer, P.R. \& Lukin, M.D. Strong coupling of single emitters to surface plasmons. \textit{Phys. Rev. B} \textbf{76}, 035420 (2007).
%
\bibitem{Tanji11}  Tanji-Suzuki, H. et al. Interaction between Atomic Ensembles and Optical Resonatots: Classical description. \textit{Adv.At.Mol.Opt.Phys.} \textbf{60}, 201 (2011).
%
\bibitem{CohenTannoudji92}  Cohen-Tannoudji, C., Dupont-Roc, J. \& Grynberg, G. \textit{Atom-photon interactions, basic processes and applications} (Wiley, New York, 1992).
%
\bibitem{Fano61} Fano, U. Effects of configuration interaction on intensities and phase shifts. \textit{Phys. Rev.} \textbf{124}, 1866–1878 (1961).
%
\bibitem{Fan14} Fan, P., Yu, Z., Fan, S. \& Brongersma, M.L. Optical Fano resonance of an individual semiconductor nanostructure. \textit{Nature Mater.} \textbf{13}, 471–475 (2014).
%
%\bibitem{Collot93} Collor, J., Lef\`evre-Seguin, V., Brune, M., \& Haroche, S. Very High-Q Whispering-Gallery Mode Resonances Observed on Fused Silica Microspheres. \textit{Europhys. Lett.} \textbf{23}, 327-334 (1993).
%\bibitem{Arc10}  Archambault, A., Marquier, F., Greffet, J.J. \& Arnold, C. Quantum theory of spantaneous and stimulated emission of surface plasmons. \textit{Phys. Rev. B} \textbf{82}, 035411 (2010).
%
%\bibitem{Gehr10}  Gehr, R. et al. Cavity-Based Single Atom Preparation and High-Fidelity Hyperfine State Readout. \textit{Phys. Rev. Lett} \textbf{104}, 203602 (2010).
%
\bibitem{Choquette10}  Choquette, J.J., Marzlin, K.-P. \& Sanders, B.C. Superradiance, subradiance, and suppressed superradiance of dipoles near a metal interface \textit{Phys. Rev. A} \textbf{82}, 023827 (2010).
\bibitem{Stratton41} Stratton, J.A. \textit{Electromagnetic Theory, Chapter 9.8} (McGraw-Hill Book Company, 1941).
\bibitem{Arc09}  Archambault, A., Teperik, T.V. , Marquier, F. \& Greffet, J.J. Surface plasmon Fourier optics. \textit{Phys. Rev. B} \textbf{79}, 195414 (2009).
%\bibitem{Vernooy98} Vernooy, D.W., Furusawa, A., Georgiades, N.Ph.,Ilchenko, V.S., \& Kimble, H.J. Cavity QED with high-Q whispering gallery modes. \textit{Phys. Rev. A} \textbf{57}, R2293 (1998).
%\bibitem{Tudela13}  Gonz\'{a}lez-Tudela, A., Huidobro, P.A., Mart\'{i}n-Moreno, C. \& Garc\'{i}a-Vidal, F.J. Theory of Strong Coupling between Quantum Emitters and Propagating Surface Plasmons. \textit{Phys. Rev. Lett.} \textbf{110}, 126801 (2013).
\bibitem{Bender10}  Bender, H. et al. Cooperative Scattering by Cold Atoms. \textit{J. Mod. Opt.} \textbf{57}, 1841-1848 (2010).
\bibitem{CT92}  Cohen-Tannoudji, C., Dupont-Roc, J. \& Grynberg, G. \textit{Atom-photon interactions, basic processes and applications} (Wiley, New York, 1992).

\end{thebibliography}
\end{document}